\documentclass[aps,prd,amssymb,superscriptaddress,showpacs]{revtex4}
\usepackage{graphicx,bm,psfrag,hyperref}
\usepackage{amsmath}
\usepackage{amssymb}
\usepackage{amsfonts}

\begin{document}

\title{QCD nature of dark energy at finite temperature: cosmological implications}

\author{K. Azizi}
\email{kazizi@dogus.edu.tr}
\affiliation{Department of Physics, Do\u gu\c s University, Ac{\i}badem Kad{\i}k\"oy, 34722 Istanbul, Turkey}

\author{N. Kat{\i}rc{\i}}
\email{nihan.katirci@boun.edu.tr}
\affiliation{Department of Physics, Bo\u{g}azi\c{c}i University, 34342 Bebek, Istanbul, Turkey and \\ Feza G\" ursey Center for Physics and Mathematics, Bo\u{g}azi\c{c}i University, 34684 \c Cengelk\" oy, Istanbul, Turkey}

\date{\today}

\begin{abstract}
The Veneziano ghost field has been proposed as an alternative source of dark energy whose energy density is consistent with the cosmological observations.
 In this model, the energy density of QCD ghost field is expressed in terms of QCD degrees of freedom at zero temperature. We extend this model to finite temperature to search
 the model predictions from  late time to  early universe. We depict the variations of QCD parameters entering the calculations, dark energy density, equation of state, 
Hubble and deceleration parameters on temperature from zero to a critical temperature. We compare our results with the observations and  theoretical predictions existing at different eras.
 It is found that this model safely defines the universe from quark condensation up to now and its predictions are not in tension with those of the standard cosmology. 
The EoS parameter of dark energy is dynamical and evolves from  $-1/3$
 in the presence of radiation to $-1$ at late time.  The finite temperature
 ghost dark energy predictions on the Hubble parameter well fit to those of $\Lambda$CDM and observations at late time.
\end{abstract}
\maketitle

\section{Introduction}
The universe is expanding at an increasing rate supported by various observations such as Supernova type 1A explosions, cosmic microwave background (CMB) radiation and baryon acoustic oscillations
 (BAO)\cite{perlmutter99,riess98,riess04,kowalski08,komatsu09,spergel06,spergel07,tegmark04}. There is a need for a kind of energy to fill roughly $75\%$ of the universe that causes the late 
time accelerated expansion. The $\Lambda$-cold dark matter (CDM) model is currently the best cosmological model explaining this expansion. It is assumed that the cosmological constant, $\Lambda$, 
may arise from vacuum fluctuations. However, there is a large (at least sixty orders of magnitude) discrepancy between the predicted energy density of the vacuum in particle physics (of the order 
 $M_p^4$ with $M_p=\frac{1}{\sqrt{G}} \approx 10^{19}~GeV$  being the Planck mass) and the energy density of the cosmological constant obtained from fitting the $\Lambda$CDM model 
predictions to observations, i.e.  $\rho_{\Lambda}^{observed}=(2.3 \times 10^{-3}~eV)^4$ \cite{perlmutter99,riess04,kowalski08,komatsu09}. This is called the cosmological constant
 problem and many models have been proposed to overcome this problem. 
One of those that relates the vacuum energy to the QCD vacuum has been proposed by F. R. Urban and  A. R. Zhitnitsky \cite{urban09,urban09-2,urban09-3,urban10} at which Veneziano ghost field 
was firstly considered as a candidate  for the late time acceleration. Veneziano first proposed this field as a ghost by putting a minus sign into the propagator with the aim of solving the
  $U(1)_A$ problem.  This field is called  ghost, since it describes the long range interactions of QCD \cite{veneziano79} (for a review see Ref. \cite{shore07}).
 In the model proposed by Urban and Zhitnitsky, the QCD vacuum energy is related to the fundamental QCD parameters and it is in the order
 of $\rho_{vac} \sim H m_q  \langle \bar{q}q \rangle/m_{\eta^{\prime}} \sim (4.3\times 10^{-3}~eV)^4$ with $H, m_q, \langle \bar{q}q \rangle$ and $m_{\eta^{\prime}} $ being 
the Hubble parameter, light quark mass, light quark condensate and mass of the $\eta^{\prime}$ meson, respectively. This energy density is in the same order of magnitude with the 
observations implying that the vacuum energy can be considered as QCD vacuum. The idea of using QCD in cosmology is not new; such that ample knowledge of QCD at
 finite baryon density and temperature has been used to understand a wide range of phenomena in cosmology  such as the structure of neutron stars,  Big Bang Nucleosynthesis (BBN)
 and so on. In Ref. \cite{schwatz98}, it is shown that different phases of QCD at finite temperature and density lead to interesting effects. In Ref. \cite{urban09-0}, it is stated that the 
gravity may be a low energy effective theory of QCD, instead of being a fundamental interaction. It is believed that at very high temperatures (higher than the critical temperature),
 the quark condensates are not  formed  and the universe consists of quark-gluon plasma (QGP). In Ref. \cite{sanches15} the equation of state (EoS) parameter of QGP is investigated.

 The standard calculations in QCD are generally performed in Minkowski space-time with topological susceptibility $\chi=0$ for zero scalar curvature. In Refs. \cite{urban09,urban09-2,urban09-3,urban10}, 
to investigate the QCD nature of dark energy, it is realized that if our universe is embedded on a nontrivial finite manifold, such as a torus, the energy density should be proportional to the 
deviation from the Minkowski space-time.
 In this model, the topological susceptibility $\chi$ is defined as
 \begin{eqnarray}
\chi \equiv i \int d^4 x e^{iqx} \langle0\mid T\{Q(x),Q(0)\} \mid0\rangle,
\end{eqnarray}
 where $Q=\partial_{\mu}K^{\mu}$ is the topological charge density and
  \begin{eqnarray}
K^{\mu}=\frac{g^2}{16\pi^2}\epsilon^{\mu\nu\lambda\sigma}A_{\nu}^{a}\left(\partial_{\lambda}A_{\sigma}^{a}+\frac{g}{3}f^{abc}A_{\lambda}^{b}A_{\sigma}^{c}\right),
\end{eqnarray}
 with $A^{a}_{\mu}$ being the conventional QCD gluon fields and $g=\sqrt{4\pi\alpha_s}$. Although $Q$ is a total derivative, the standard Witten-Veneziano solution of $U(1)_A$ problem
 implies that $\chi$ does not vanish. This causes an unphysical pole at zero momentum in the correlation function of $K_{\mu}$. Using $\langle 0 \mid K_{\mu} \mid ghost \rangle=\lambda_{YM} \epsilon_{\mu}$; 
with $\lambda_{YM}$ and $\epsilon_{\mu}$ being the decay constant and   the polarization vector, respectively; as well as  $\sum \epsilon_{\mu} \epsilon_{\nu}=-g_{\mu\nu}$, one can write
  \begin{equation}
  \lim_{q\to0} i \int d^4 x e^{iqx} \langle0\mid T\{Q(x),Q(0)\} \mid0\rangle =-\frac{g^{\mu\nu}}{q^2} \lambda_{YM} q_{\mu}q_{\nu},
  \label{corre}
\end{equation}
 where $\frac{i g^{\mu\nu}}{q^2}$ is the propagator of ghost field and the minus sign is considered to solve the  $U(1)_A$ problem \cite{urban09}. By introducing a single light quark with mass $m_q$ 
and the matrix element of the $\eta^{\prime}$ field via $\langle0 \mid K_{\mu}\mid \eta^{\prime}\rangle=\frac{i \lambda_{\eta^{\prime}}}{\sqrt{N_c}}q_{\mu}$ one gets
  \begin{equation}
 \chi= \lim_{q\to0} i \int d^4 x e^{iqx} \langle0\mid T\{Q(x),Q(0)\} \mid0\rangle =\frac{(q^2-m_0^2) \lambda_{YM}^2}{(q^2-m_{\eta^{\prime}}^2)},
  \label{corre2}
\end{equation}
where $m_{\eta^{\prime}}^2=m_0^2+\lambda_{\eta^{\prime}}^2/N_c$ is the mass squared of the physical $\eta^{\prime}$ field. On the other hand, the relevant Ward identity for QCD with light quarks
  \begin{equation}
 \chi \equiv i \int d^x e^{iqx} \langle0\mid T\{Q(x),Q(0)\} \mid0\rangle =m_q\langle \bar{q}q\rangle+O(m_q^2),
  \label{corre}
\end{equation}
 is satisfied.
By using these relations and $m_0^2f_{\pi}^2=-4m_q\langle \bar{q}q \rangle$, the famous Witten-Veneziano relation, $4\lambda_{YM}^2=f_{\pi}^2m_{\eta^{\prime}}^2$, is obtained.
The deviation mentioned above is related to the finite size (L) of the manifold and $\Delta\lambda_{YM}=\frac{1}{L}=H$. The topological susceptibility is related to the energy density
 through \cite{urban09,witten79}
 \begin{eqnarray}
\chi=-\frac{\partial^2\epsilon_{vac}(\theta)}{\partial \theta}\mid_{\theta=0},
\label{energydensity}
\end{eqnarray}
where $\theta$ is the angle of QCD. The corrections due to  very large but finite size of the manifold are small. It is important to mention that if one knows the $\theta$ dependence of vacuum energy, 
one can compute the energy mismatch that arises in theory between the infinite Minkowski and finite compact space-times. This effect is entirely due to the ghost and only much smaller corrections arise 
from all other QCD fields.
From  \eqref{corre2}-\eqref{energydensity}, one can write (for details, see Ref. \cite{urban09})
\begin{equation}
\Delta\left[\frac{\partial^2\epsilon_{vac}(\theta)}{\partial \theta^2}\mid_{\theta=0}\right]=-\Delta \chi = \Delta \left(\frac{m_0^2 \lambda_{YM}^2} {m_{\eta^{\prime}}^2} \right) \simeq -c \frac{2H}{m_{\eta{\prime}}}  \left(\frac{m_0^2 \lambda_{YM}^2} {m_{\eta^{\prime}}^2} \right)\simeq  -c \frac{2N_fH}{m_{\eta{\prime}}}\mid m_q \langle \bar{q}q \rangle\mid < 0.
\label{final}
 \end{equation}
 The $\theta$ dependence of vacuum energy at $\theta\ll1$ and for $N_f$ quark flavors with equal mass is given as  $\epsilon_{vac}=-N_f\mid m_q \langle \bar{q}q \rangle\mid cos{\frac{\theta}{N_f}}$ in Ref. \cite{urban09,witten80,vecchia80}, such that $\partial^2 \epsilon_{vac}(\theta)=-\frac{\epsilon_{vac}}{N_f}$.
 Note that all contributions from the gluon condensates and condensates from the Higgs field and etc., cancel out in the subtraction as they appear with equal magnitude. From \eqref{final} one can write
 \begin{equation}
\rho_g \equiv \Delta \epsilon_{vac}=c \frac{2N_fH}{m_{\eta{\prime}}} \mid m_q \langle \bar{q}q \rangle\mid.
 \end{equation}

This energy density in an expanding background, the  Friedmann-Robertson-Walker (FRW) space-time, is analyzed in Ref. \cite{urban10}. The global topology could be a torus and 
 FRW metric still locally describes the space-time. The universe may have a non-trivial topology and there are different searches on this, such as 
  the matched circle test.  It must be mentioned that 
these searches yield no detection of a compact topology \cite{spergel12,aslanyan12,planck15}. 

 In Ref. \cite{urban10}, Urban and Zhitnitsky use the QCD vacuum energy in  FRW space-time to calculate some cosmological parameters and compare the model predictions with those of the $\Lambda$CDM and 
cosmological observations. In the calculations, they consider the QCD parameters at zero temperature (late time) to investigate the evolutions of EoS parameter of dark energy and Hubble parameter
 at low redshifts. In the present study, we extend those calculations by considering the QCD parameters at finite temperature. By increasing the temperature we can go from  late time to the early universe 
and look for the variations of QCD vacuum, and as a result, variations of the cosmological parameters with respect to time (temperature).
 In particular, we use the temperature dependent $m_{\eta^{\prime}}(T)$ and $\langle\bar{q}q\rangle(T)$ to investigate the ghost energy density parameter ($\Omega_g$), EoS parameter
 of ghost dark energy ($\omega_g$), total EoS parameter ($\omega_{tot}$), Hubble ($H$), and deceleration ($q$) parameters as a function of e-folding N and  redshift z. 
The results are compared with  those of zero temperature existing in the literature as well as the $\Lambda$CDM.

The outline of the paper is as follows. We present some details of the model at finite temperature and its modifications on the energy conservation in the following Section.
 In Section III, we discuss the cosmological parameters and their time evolution. The last section is devoted to the concluding remarks.

\section{The ghost energy density at finite temperature}
As previously discussed, the energy density of QCD ghost field, $\rho_{g}$, can be related to the Hubble parameter at late time as
\begin{eqnarray}
\rho_{g}=\alpha_0 H_0,
\label{rhoghost}
\end{eqnarray}
with
\begin{eqnarray}
\alpha_0=\frac{2cN_F | m_q \langle \bar{q}q \rangle_0|}{m_{\eta^{\prime}}},
\label{alpha}
\end{eqnarray}
where $\langle \bar{q}q\rangle_0=(-240 \times 10^6 ~eV)^3$ is the light quark condensate at zero temperature, $m_q=3.5\times10^{-3}~MeV$ 
 is the average light quark mass (up and down), $m_{\eta^{\prime}}=957.78\pm0.06~MeV$  \cite{pdg14} is the mass of $\eta^{\prime}$ meson
 at vacuum, $c$ is the speed of light, $H_0$ is today's value of the Hubble parameter ($H=\frac{\dot{a}}{a}$) and $N_F$ denotes the number of flavors.
 We  use the natural units $8\pi G=\frac{8\pi}{M_P^2}=c=1$.

The parameter $\alpha_0$ is constant at zero temperature. By setting $N_F=2$ and putting all other values, the energy density of QCD ghost field  is found as $\rho_g \sim (4.3\times 10^{-3}~eV)^4$
 which leads to a consistent prediction with the observations at late time. To investigate the dependence of $\alpha$ on temperature, we need to know the dependence of the quark 
condensate $\langle \bar{q}q\rangle$ and $m_{\eta^{\prime}}$ on temperature. In the present study, we use the parametrization for the behavior of the quark condensate in terms 
of the temperature given in Ref. \cite{ayala11} which can be fitted to the following function:
\begin{eqnarray}
\langle \bar{q}q\rangle (T)=\langle \bar{q}q\rangle_0\left(\frac{1}{1+e^{18.10042(1.84692T^2+4.99216T-1)}})\right)~(GeV^3).
\label{qqbar}
\end{eqnarray}
This parametrization reproduces the lattice QCD predictions on the temperature-dependent light quark condensates \cite{Bazavov09,Cheng10}.

The behavior of $\eta^{\prime}$ meson mass with respect to temperature is studied in Ref. \cite{hioki97} and can be parametrized in terms of temperature as
\begin{eqnarray}
m_{\eta^{\prime}}(T)=0.958- 0.082T + 6.127T^2 - 79.287 T^3~(GeV),
\label{etaprime}
\end{eqnarray}
where we took the critical temperature $T_c=197~MeV$  (see also Ref. \cite{ayala12}).
Using the above parametrizations, we plot the dependence of the quark condensate, $m_{\eta^{\prime}}(T)$, and $\alpha(T)$ in terms of temperature up to $T=220~MeV$  in figure 1. From this 
figure we see that $\langle \bar{q}q\rangle$ and $m_{\eta^{\prime}}$ are constant up to some temperatures after which they start to diminish, drastically.
 The parameter $\alpha$ also remains unchanged  up to roughly $0.16~GeV$,  after which it immediately falls to zero.
\begin{figure}[h!]
\centering
\begin{tabular}{ccc}
\includegraphics[width=5cm,height=4cm]{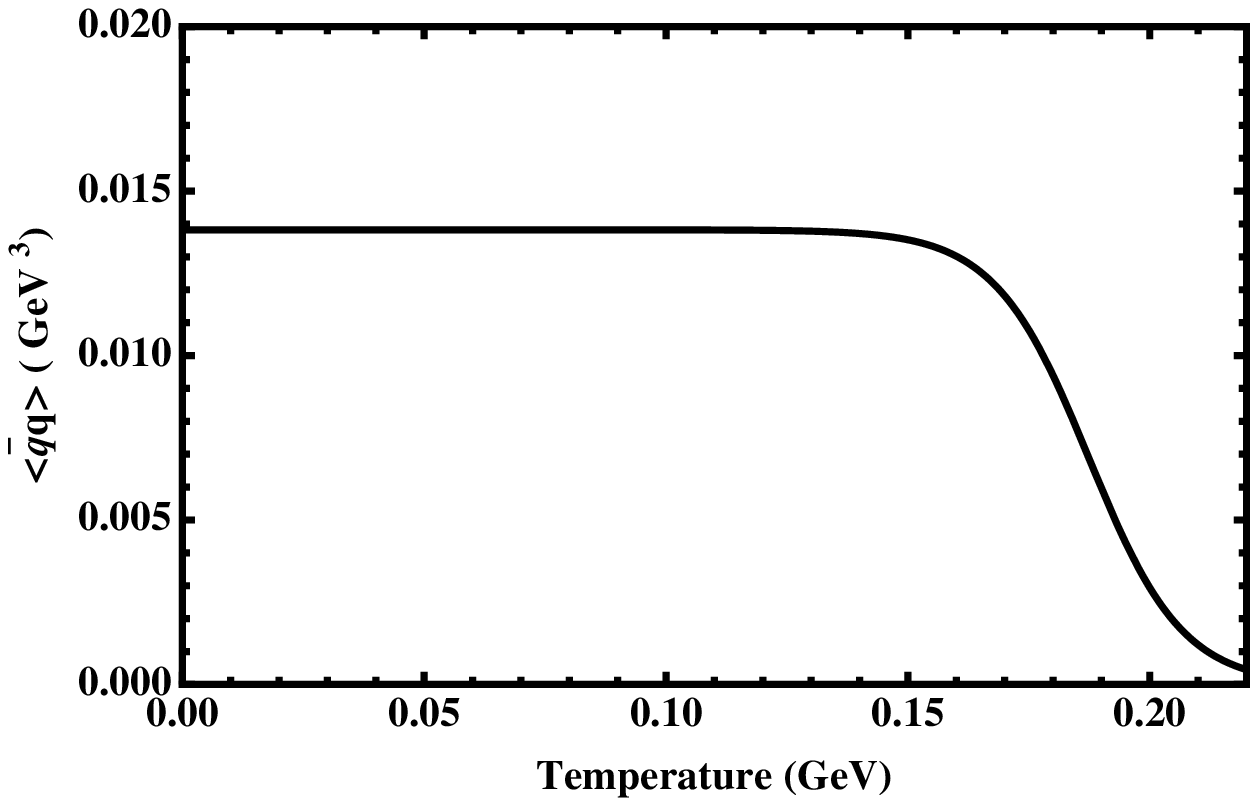}
\includegraphics[width=5cm,height=4cm]{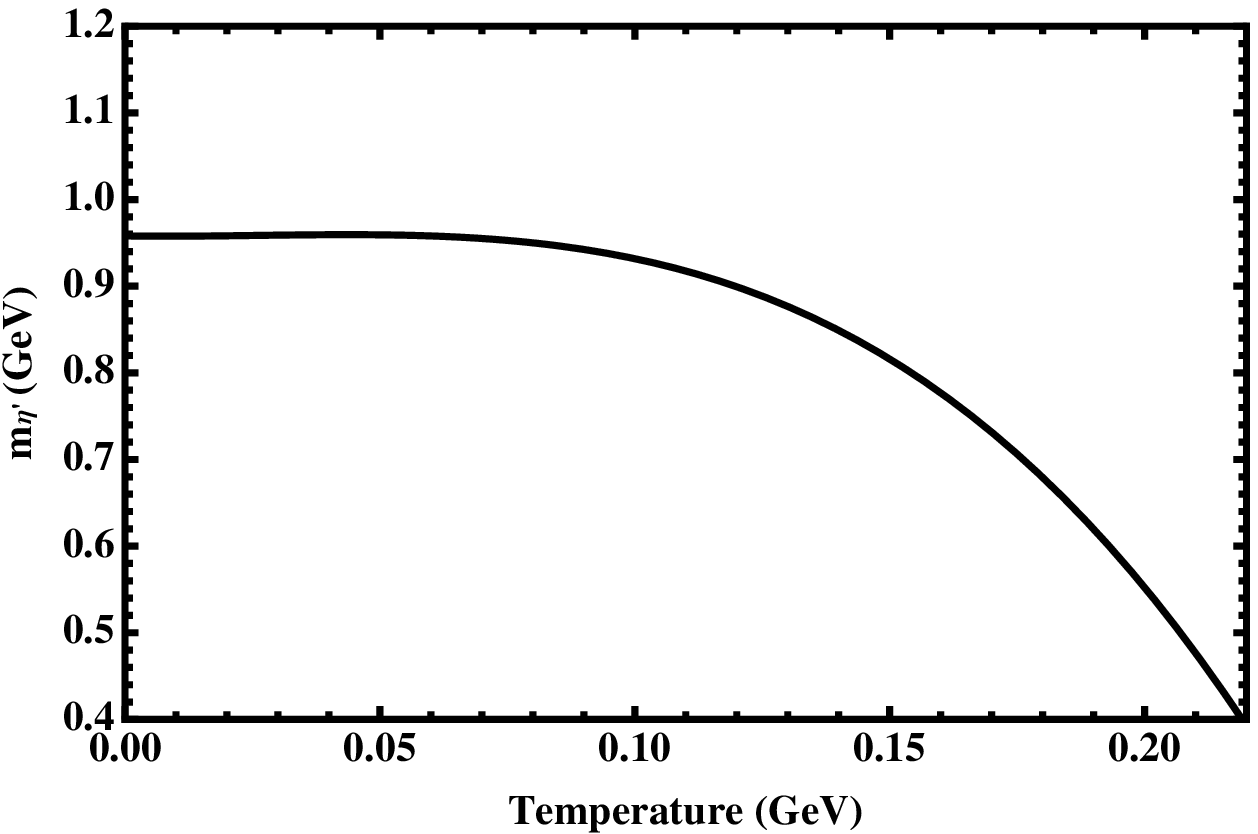}
\includegraphics[width=5cm,height=4cm]{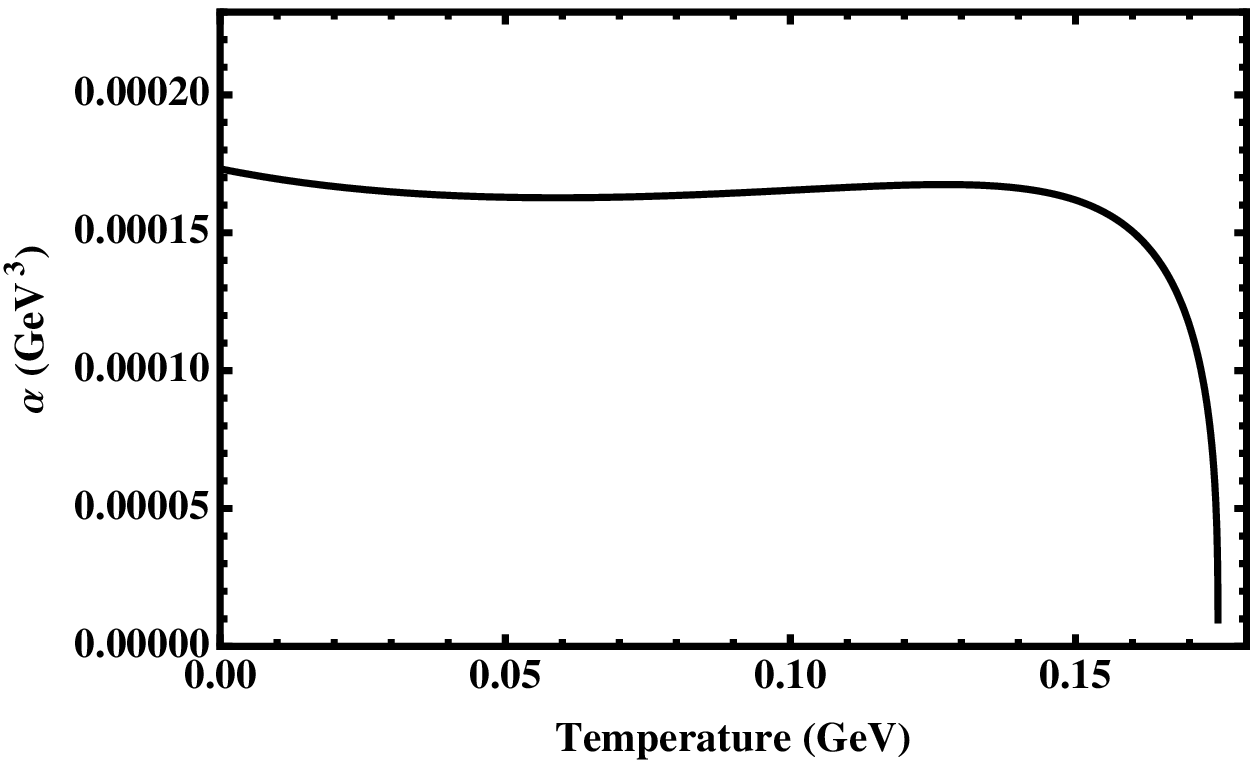}
\end{tabular}
\label{temp}
\caption{The dependence of  $\langle \bar{q}q\rangle $, $m_{\eta^{\prime}}$ and $\alpha$ on temperature.}
\end{figure}
We shall remark that the $\alpha$ parameter is  barely constant for the late time as  seen from the figure.
To simplify the calculations, we use the e-folding (redshift) instead of temperature in the following. The relation between the scale factor, redshift and e-folding is 
\begin{equation}
\eta \frac{T}{T_0}=\frac{a_0}{a}=1+z=\frac{e^{N_0}}{e^N},
\label{rel}
\end{equation}
where $N=\ln{a}$ and subindex $``0"$  stands for today's values. The parameter  $\eta$ takes the values $\eta=\left(\frac{11}{4}\right)^{\frac{1}{3}}$ for before 
the electron formation ($T>m_e$) and $\eta=1$ for the first electron formation up to now ($T<m_e$). The dependence of the parameters under consideration on  
N/z is shown in figure 2. From this figure it is clear that the $\alpha$ remains unchanged up to $N=-26.5$, after which it starts to grow then immediately falls near to $N=-28$.
\begin{figure}[h!]
\centering
\begin{tabular}{ccc}
\includegraphics[width=5cm,height=4cm]{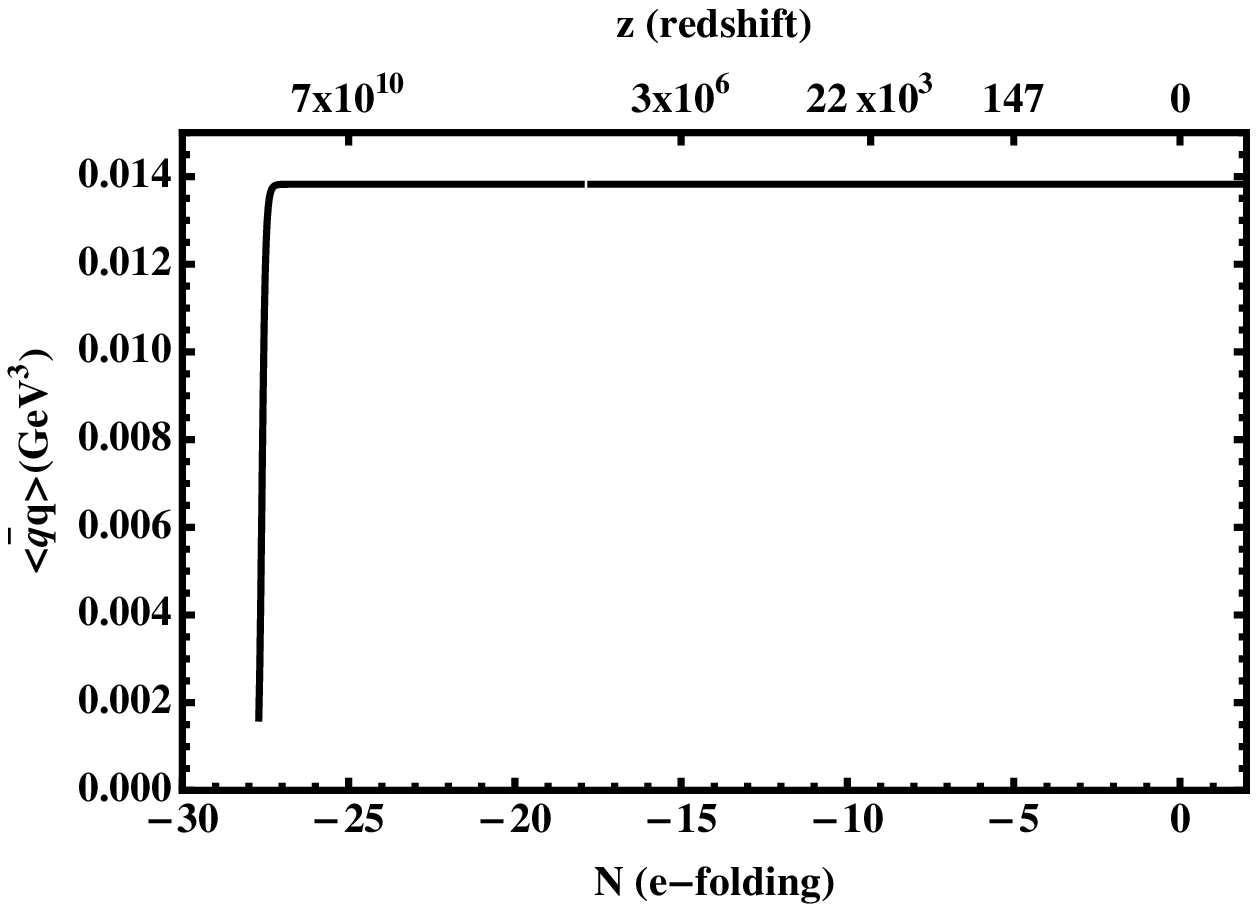}
\includegraphics[width=5cm,height=4cm]{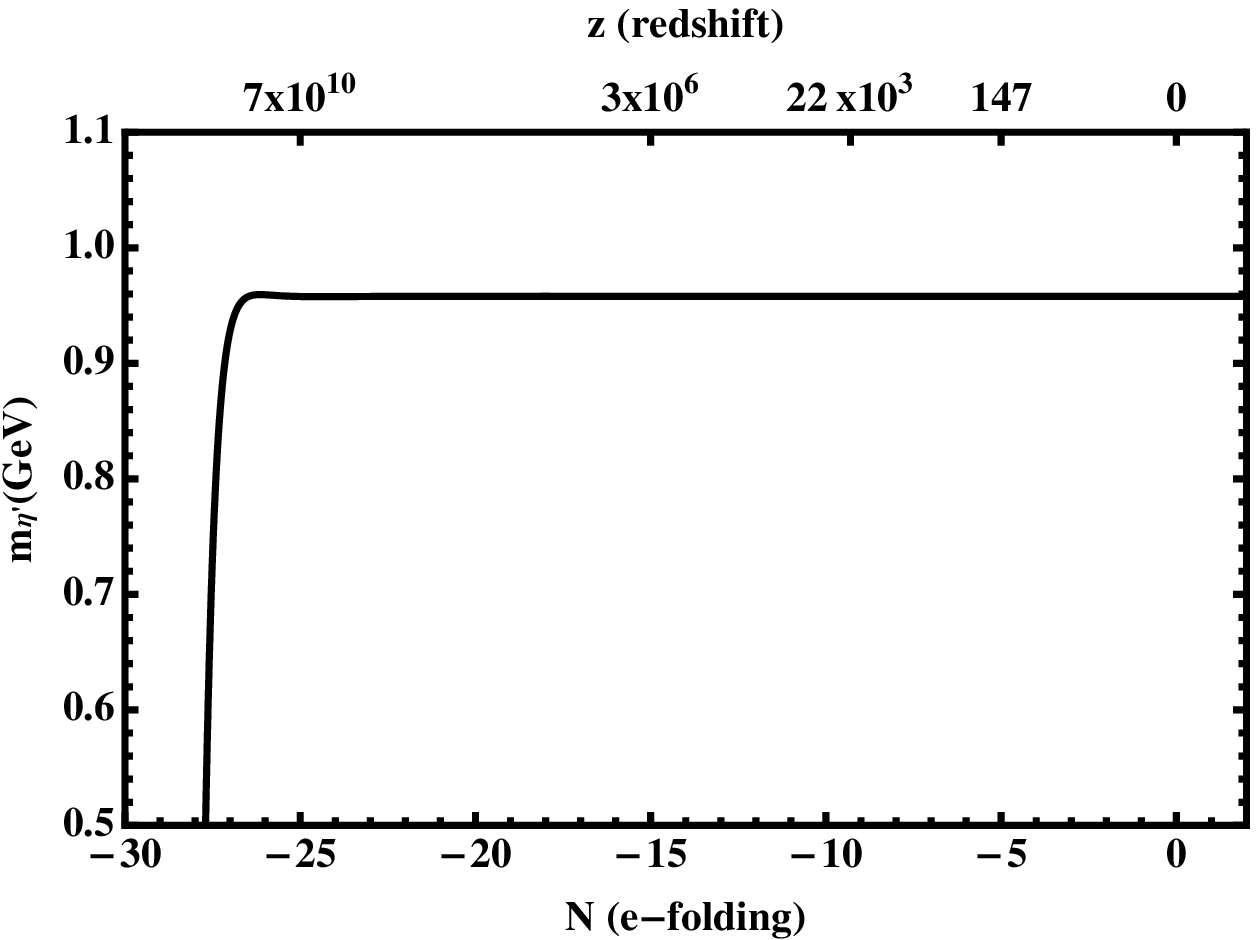}
\includegraphics[width=5cm,height=4cm]{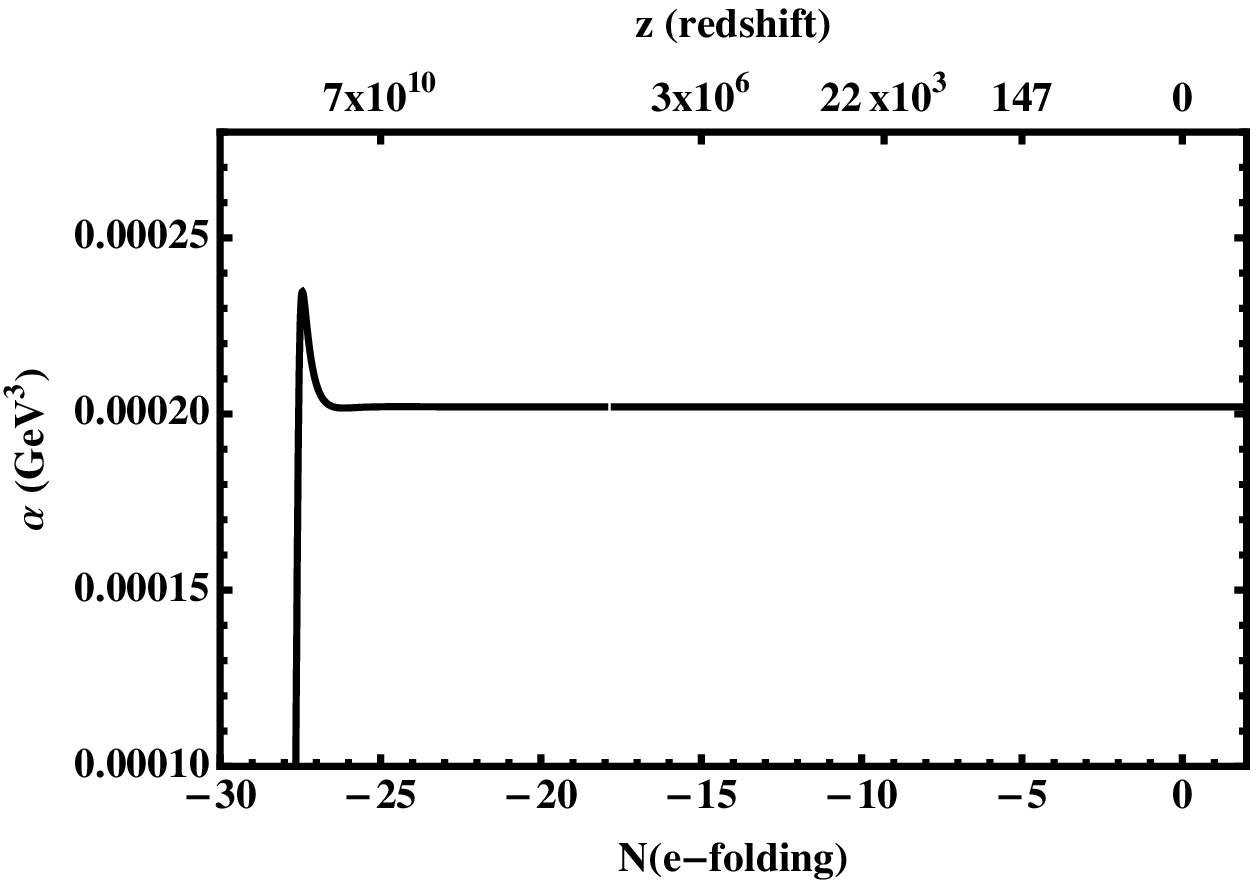}
\end{tabular}
\label{fig:etaqqbar}
\caption{The $\langle \bar{q}q\rangle$, $m_{\eta^{\prime}}$ and $\alpha$ as a function of N (e-folding)/z (redshift).}
\end{figure}

Before delving into details, we would like to overview the constant QCD ghost implications in cosmology and compare it with the temperature dependent QCD vacuum. 
The Friedmann equation
   \begin{equation}
 3H^2M_p^2= \alpha_0 H,
\end{equation}
gives the  exact solution of the scale factor for constant $\alpha$ as
 \begin{equation}
  a(t)=a_0 e^{\frac{\alpha_0}{3M_p^2}t}.
  \end{equation}
The constant $\alpha$ at late time also implies $w_g=-1$ for the ghost dark energy and this gives the de-Sitter type expansion. When we take the temperature into account,
 the solution of Friedmann equation becomes more complicated.
The continuity equation
\begin{eqnarray}
\dot{\rho}_{g}+3(1+w_{g})H\rho_{g}=0,
\label{cont}
\end{eqnarray}
is naturally modified to
 \begin{eqnarray}
(\alpha \dot{H}+\dot{\alpha}H)+3(1+w_g)H\rho_{g} & = &0,\label{noncons}
\end{eqnarray}
at finite temperature. The extra $\dot{\alpha}H$ term may be considered as an interaction term between the ghost field and other components of the universe. 
The interaction type is determined by the temperature dependence of QCD parameters but this is not investigated in the present paper.
 The cosmic evolution of the universe  in chronological order from Big Bang up to now   is inflation, radiation dominated era, matter dominated era, recombination,
 structure formation and late time accelerated expansion. In this paper, from now till the radiation-matter equality ($T \approx 3 \times 10^4$ K, t $\approx 10^4$ yr, $z_{rm}=2740$, $N_{rm}=-9.3$), 
the QCD ghost field and matter are considered  to fill the universe and the radiation is safely neglected. After the radiation-matter equality, the radiation dominates over  matter and the QCD vacuum  
 and radiation are considered to fill the universe. 
\section{ Analysis of Some Cosmological parameters at finite temperature}
\subsection{Energy density parameter of ghost dark energy}
At late time  (the ghost field and  matter dominated era), the Friedmann equation is written as
\begin{eqnarray}
3H^2M_p^2=\rho_{g}+\rho_{m},
\label{fried}
\end{eqnarray}
which satisfies  the continuity equations
\begin{eqnarray}
\dot{\rho}_{g}+3\left(1+w_{g}\right)H\rho_{g}=0 , ~~~~\mbox{and}  ~~~\dot{\rho}_{m}+3H\rho_{m}=0.
\label{continuities}
\end{eqnarray}
Dividing both sides of \eqref{fried} by the critical energy density $\rho_c=3H^2M_p^2$, we obtain  $\Omega_{g}+\Omega_{m}=1$ where $\Omega_x=\frac{\rho_x}{\rho_c}$ is the dimensionless energy 
density parameter of $x$. The continuity equation for matter can be written in terms of  $\Omega_m$ as
\begin{equation}
\dot{\Omega}_m+\left(2\frac{\dot{H}}{H}+3H\right)\Omega_m=0,
\label{rhodot}
\end{equation}
and in terms of $\Omega_{g}$ as
\begin{equation}
-\dot{\Omega}_g+\left(2\frac{\dot{H}}{H}+3H\right)(1-\Omega_g)=0.
\label{omegag}
\end{equation}
The rate of change in Hubble parameter  is related to $\Omega_g$ as
\begin{eqnarray}
\frac{\dot{H}}{H}= \frac{\dot{\alpha}}{\alpha}-\frac{\dot{\Omega}_{g}}{\Omega_{g}},
\label{hdot}
\end{eqnarray}
hence, \eqref{omegag} becomes
 \begin{eqnarray}
\dot{\Omega}_g(\Omega_g-2)+\left(2\frac{\dot{\alpha}}{\alpha}+3H\right)(1-\Omega_g)\Omega_g=0.
\label{onlyomega}
\end{eqnarray}

Let us rewrite \eqref{onlyomega} in terms of e-folding  as
 \begin{equation}
\Omega_g'(\Omega_g-2)+\left(2\frac{\alpha'}{\alpha}+3\right)(1-\Omega_g)\Omega_g=0,
\label{onlyomegaN}
\end{equation}
where $'$ denotes the derivative with respect to  e-folding.  

For  radiation and  ghost field dominated era, considering the continuity equation for radiation,
 $\dot{\rho}_{r}+4H\rho_{r}=0$, from a similar manner, we obtain
 \begin{equation}
\Omega_g'(\Omega_g-2)+(2\frac{\alpha'}{\alpha}+4)(1-\Omega_g)\Omega_g=0.
\label{onlyomegaNrad}
\end{equation}

\begin{figure}[h!]
\begin{centering}
\includegraphics[width=7cm,height=5cm]{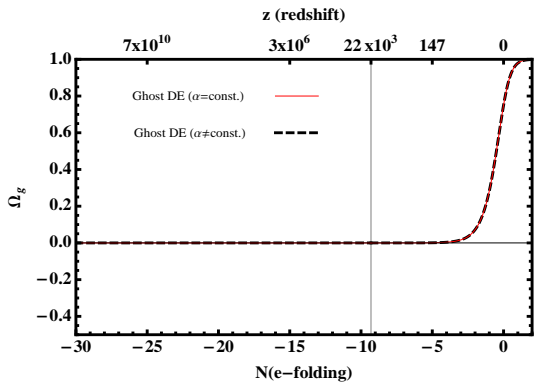}
\caption{The dependence of ghost dark energy density parameter $\Omega_g$ on N/z.}
\label{fig:omradmat}
\end{centering}
\end{figure}

%In this paper, the two regions are analyzed independently but the graphs are plotted to contain both eras together. 
The $\Omega_g$ versus $N$ for the two eras,
 obtained by the numerical solving of  \eqref{onlyomegaN} and \eqref{onlyomegaNrad}, is depicted in figure \ref{fig:omradmat}. 
The vertical line on this graph shows the radiation-matter equality.
 From figure \ref{fig:omradmat}, we see that the ghost dark energy density at $T\neq0$ evolves exactly the same as the zero temperature case ($\alpha=const.$). 
From this figure, we also see that there is no role of the ghost dark energy for  $N<-3$ and it is produced when $N>-3$ in the universe. 
In Ref. \cite{cai11}, the $\Omega_{g}$ is analyzed in the interval $-4 < N < 2$ at zero temperature  by considering no interaction between the ghost dark energy and CDM.
 We extend the interval up to $N \simeq-30$. Our result ($\alpha \neq const.$) is in a good agreement with the predictions of Ref. \cite{cai11} for $\Omega_g$ versus $N$ in the interval $-4 < N < 2$.

\subsection{The EoS parameter of ghost dark energy}
\begin{figure}[h!]
\begin{centering}
\begin{tabular}{cc}
\includegraphics[width=7cm,height=5cm]{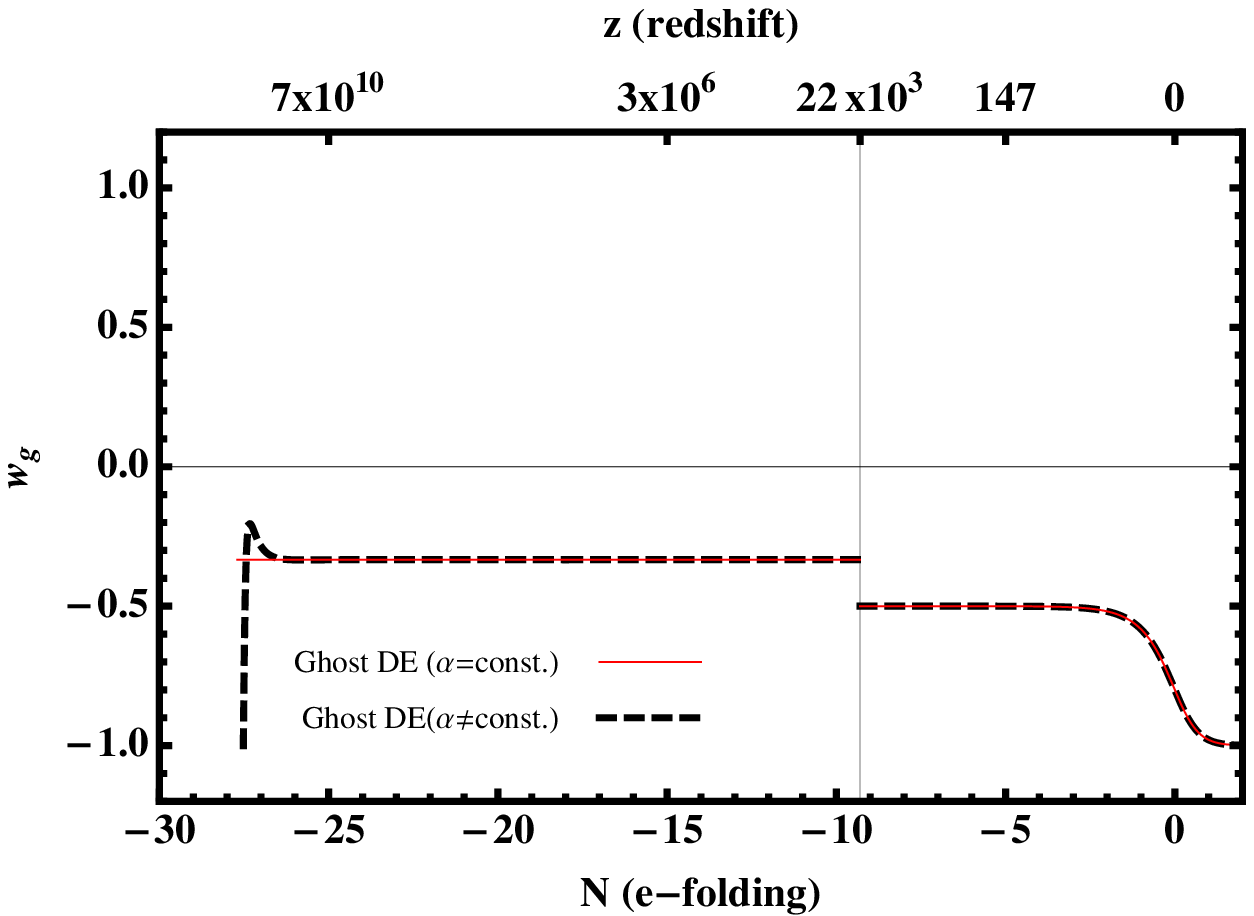}
\includegraphics[width=7cm,height=5cm]{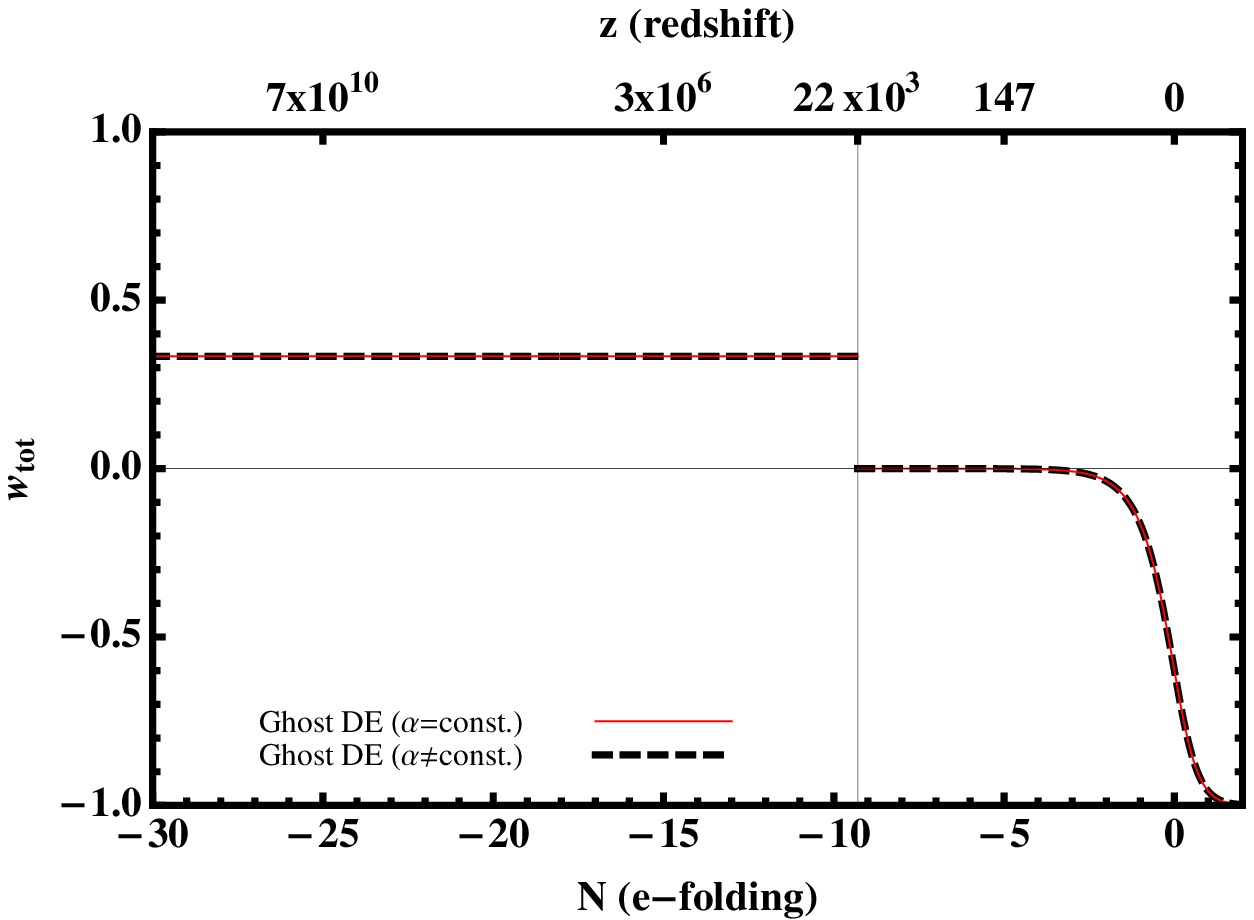}
\end{tabular}
\caption{The EoS parameter $w_{g}$ versus N/z (left panel). The total EoS parameter $w_{tot}$ versus N/z (right panel).} \label{fig:wtotal}
\end{centering}
\end{figure}

For  late time, rewriting the continuity equation for ghost dark energy \eqref{continuities} in terms of $\Omega_g$ gives
\begin{equation}
\dot{\Omega}_g+\left(2\frac{\dot{H}}{H}+3H(1+w_g)\right)\Omega_g=0.
\label{ome}
\end{equation}
Using this equation together with \eqref{omegag} and \eqref{hdot}, we get the EoS parameter as
\begin{equation}
w_{g,1}=\frac{1}{(2-\Omega_g)}\left(-1-\frac{2}{3}\frac{\alpha'}{\alpha}\right),
\label{eosmat}
\end{equation}
where the subindex $1$ refers to the late time.
Before the radiation-matter equality, in the presence of radiation and ghost dark energy, the EoS  parameter reads
\begin{equation}
w_{g,2}=-\frac{1}{3(2-\Omega_g)}\left(2\frac{\alpha'}{\alpha}+\Omega_g+2\right),
\label{eosrad}
\end{equation}
with the subindex $2$  being representing the  radiation and  ghost dark energy dominated era. We plot the dependence of $w_{g}$ on N/z for the two eras in the left panel
 of figure \ref{fig:wtotal}. We also would like to calculate the EoS parameter of the total fluid for  two above mentioned eras.
Since $w_m=0$, from the radiation-matter equality up to the finite future, $N=2$, we have
\begin{equation}
w_{tot}=w_{g,1}\Omega_g.
\label{wtotmat}
\end{equation}

From QCD phase transition ($N_{qcd} \sim -27$) up to radiation-matter equality ($N_{rm}=-9.3$), we can also  write
\begin{equation}
w_{tot}=w_{g,2}\Omega_g+(1-\Omega_g)w_r,
\label{wtotrad}
\end{equation}
where $w_r=1/3$. We  plot the dependence of $w_{tot}$ on N/z for the two eras in the right panel of figure \ref{fig:wtotal}.

 Before the description of the results presented in figure \ref{fig:wtotal} and discussion on the evolution of EoS parameter with respect to e-folding,
 we shall remind that the $\Lambda$CDM defines a flat universe with a cosmological constant ($w_{\Lambda}=-1$) and $w$CDM extends this model to allow the EoS parameter to be different than $-1$.
 The Planck collaboration gives $w =-1.13_{-0.25}^{+0.23}$ from combined Planck+WP+highL+BAO data \cite{Planck13} and BOSS collaboration gives $w=-0.97 \pm 0.05$  
from the most recent combined Planck+BAO+CMB data \cite{Boss14}. 
In figure \ref{fig:wtotal},  when going from  early to late time, we see that our model predicts $w_g=-1$ near to the critical temperature but it immediately increases to $w_g=-\frac{1}{3}$ 
and remains unchanged up to the radiation-matter equality  for $\alpha \neq const$. 
This means that the ghost dark energy behaves as a cosmic string in the presence of radiation. However, in this interval, the QCD ghost is dominated by the radiation (see the right panel) 
and thus we can safely say that the QCD ghost does not change the BBN predictions. Our results on the $w_g$ and $w_{tot}$ are consistent with predictions of Ref. \cite{cai11} for non-interacting case
 existing in the interval $-4 < N < 2$.  

After radiation matter equality, the ghost field EoS parameter becomes roughly $-\frac{1}{2}$ in the presence of matter and remains unchanged up to $N \sim -1$, after which it evolves to $-1$ for now. 
In this era, the total EoS parameter is zero referring to the deceleration in expansion up to $N \sim -3$ then it starts to drop to $-1$ representing the late time accelerated expansion. 
The transition from the deceleration ($w_{tot}>-\frac{1}{3}$)  to acceleration ($w_{tot}< -\frac{1}{3}$) in expansion occurs at $N=-0.51~(z \sim 0.6)$. This is a bit late compared to the $\Lambda$CDM prediction at which this transition takes place at $N= -0.55$ ($z \sim 0.7$).

It is interesting that  such variation of  EoS parameter for dark energy at different eras has been predicted in some  independent studies \cite{spergel99,silk04,vilenkin12,akarsu15}. 
They have found $w=-\frac{2}{3}$ for early time showing that forms of matter such as domain walls do exist, and it evolves to $w=-1$ for  late time. Similarly, in Ref. \cite{akarsu15},
 the authors  re-parameterize the dark energy source and find that the dark energy source with a dynamical EoS parameter $-\frac{2}{3}$ at early periods of universe and $-1$ today,
 matches slightly better than $\Lambda$CDM model  to the recent observations. In this work, we find that the QCD ghost dark energy with a dynamical EoS parameter starts 
from the value $-\frac{1}{3}$ at earlier times and goes to $-1$, behaving similar to the cosmological constant as time evolves.

\subsection{Hubble and deceleration parameters}
In $\Lambda$CDM model,  the Hubble parameter is identified in terms of e-folding as
\begin{equation}
\frac{H^2}{H_0^2}= \Omega_{\Lambda,0}+\Omega_{m,0} e^{-3N}+\Omega_{r,0} e^{-4N},
\label{H}
\end{equation}
in  FRW space-time for flat space-like sections. Here, the density parameters satisfy $\Omega_g+\Omega_m+\Omega_r=1$; $H_0=70.6~km/sec/Mpc$ , $\Omega_{\Lambda}=0.72, \Omega_m=0.28$ \cite{delubac14} and 
$\Omega_{r,0}=2.47 \times 10^{-5}$ are considered. The  Hubble parameter in terms of  $w_{tot}$ can be written as
\begin{equation}
\dot{H}+\frac{3}{2}H^2(1+w_{tot})=0.
\label{Hour}
\end{equation}
 \begin{figure}[h!]
\begin{centering}
\begin{tabular}{cc}
\includegraphics[width=7cm,height=5cm]{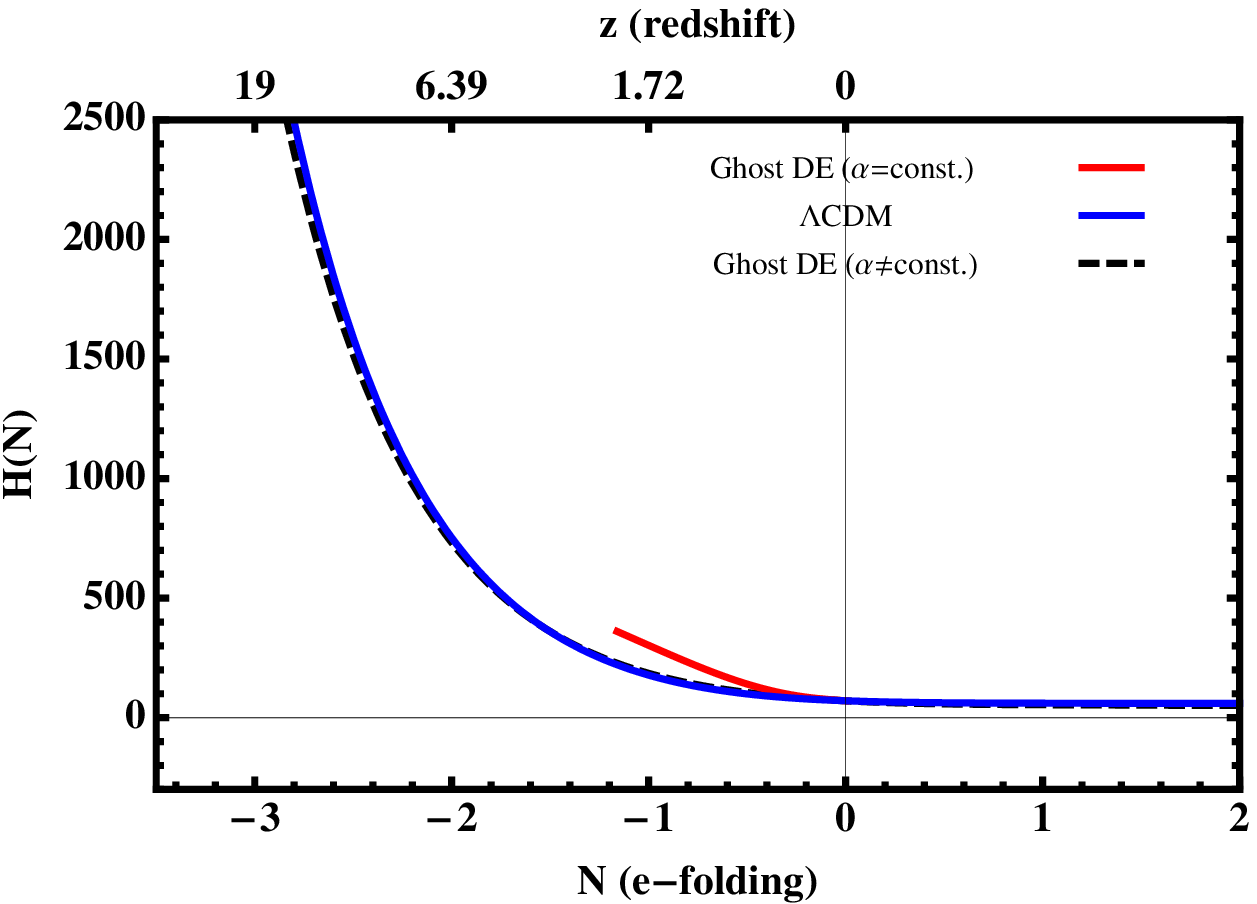}
\includegraphics[width=7cm,height=5cm]{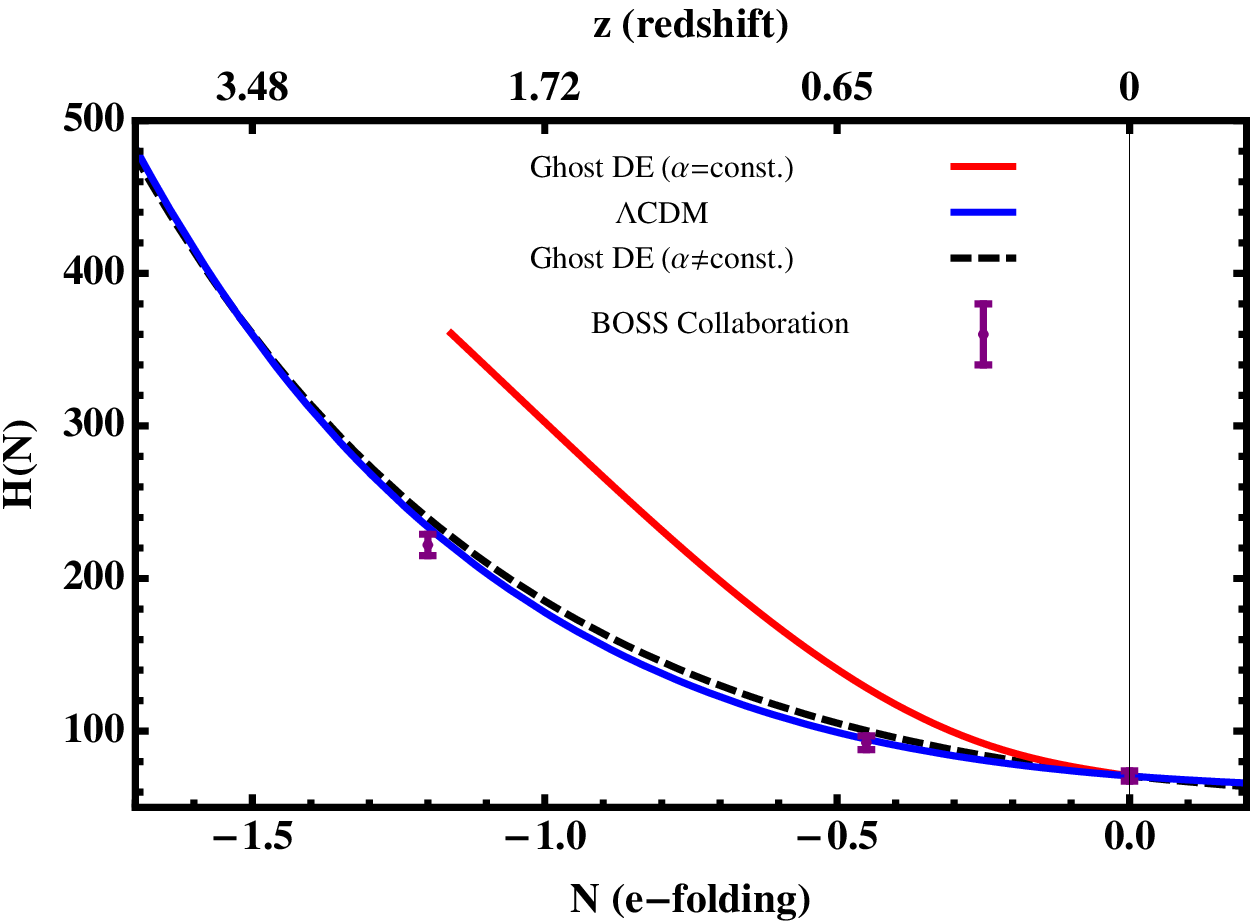}
\end{tabular}
\caption{Hubble parameter versus N/z (left panel). Zoomed version of $H$ versus N/z together with data from the BOSS Collaboration \cite{sahni14,delubac14} (right panel).} \label{fig:mathubble}
\end{centering}
\end{figure}
The variation  of Hubble parameter (left panel) and its zoomed version (right panel) in terms of N/z are depicted  in figure \ref{fig:mathubble}.
For comparison, in the same figure, we also show the variation of Hubble parameter in  $\Lambda$CDM model  as well as the variation of this parameter obtained at constant  $\alpha$   by Ref. \cite{urban10}, i.e.
\begin{equation}
\frac{H^2}{H_0^2}= \Omega_{g,0}e^{-3(1+w_g(N))N}+\Omega_{m,0} e^{-3N}+\Omega_{r,0} e^{-4N}.
\label{Hurban}
\end{equation}
 We also add the most recent data \cite{sahni14,delubac14} to figure \ref{fig:mathubble}.

From this  figure, we see that our model's prediction ($\alpha \neq const.$) on variation of $H$ is very close to that of the $\Lambda$CDM model and very well fits to the observational data. 
The behavior of $H$ versus $N$ for $\alpha=const.$ obtained in Ref. \cite{urban10} given in the interval $-1.2<N<0$ considerably differs from the others in the interval $-1.2<N<-0.2$, 
although it predicts the same $H$ with other models and observational data at $N=0$. From this figure, we also see that, when going from the late to early time, the Hubble parameter 
drastically increases in terms of e-folding.

\begin{figure}[h!]
\includegraphics[width=7cm,height=5cm]{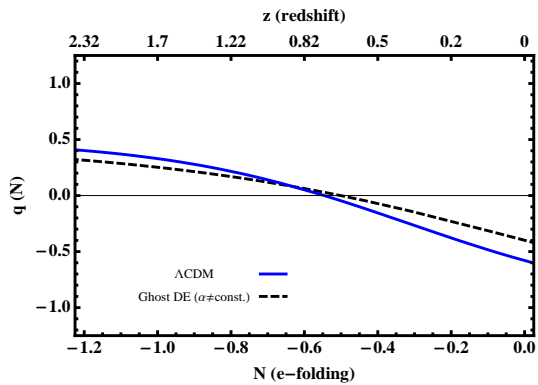}
\caption{Deceleration parameter versus N/z.}  \label{fig:matq}
\end{figure}

Another cosmological parameter is the deceleration parameter, defined as
 \begin{eqnarray}
 q=-1-\frac{\dot{H}}{H^2}=\frac{1+3w_{tot}}{2}.
\label{qrad}
\end{eqnarray}
 The deceleration parameter $q$ versus N/z for  late time is depicted in figure \ref{fig:matq}.
From this figure, we read that the behavior of $q$ with respect to $N$ at late time shows roughly good consistency with the predictions of $\Lambda$CDM. 
It is well known that the transition from positive to negative values of $q$ refers to the transition from deceleration to the present acceleration in expansion.
 As  seen from  figure \ref{fig:matq}, in the  $\Lambda$CDM model, this transition occurs at $N \sim -0.55$ ($z \sim 0.7$), while in our model, 
the acceleration starts a bit later at nearly $N\sim -0.51 (z \sim 0.6 )$.

\section{Conclusions}
The ghost dark energy model proposed by Urban and Zhitnitsky \cite{urban09,urban09-2,urban09-3,urban10} at zero temperature gives the energy density compatible with the 
cosmological observations at late time. We extended their model from late to early periods of the universe and searched for the variations of some cosmological observables in terms of
 temperature  by considering the temperature-dependent  QCD parameters. We discussed the variations of  the energy density parameter of ghost dark energy, the EoS parameters of the ghost and total 
fluid, the Hubble and deceleration parameters with respect to N/z.

It has been found that the ghost dark energy plays no role at early universe and it is produced after $N> -3$. The lack of  ghost dark energy before $N> -3$ is in agreement
with the predictions of the standard Big Bang on the decelerated expansionary phases  (matter and radiation dominated eras).   Our prediction on $\Omega_g$ is in a good consistency
with the prediction of Ref. \cite{cai11} existing in the interval $-4<N<2$ for constant $\alpha$ and non-interacting case.

The model predicts the EoS parameter of ghost dark energy to be   $w_g=-1$ near to the critical temperature, although it drastically increases to $w_g=-\frac{1}{3}$
referring to the behavior as a cosmic string in the presence of radiation. It  remains unchanged up to the radiation matter equality. 
In this interval, the total EoS parameter of the fluid shows that the QCD ghost is dominated by the radiation and the predictions stay consistent with the BBN's and so with the values for the abundance 
of light elements. After radiation matter equality, the ghost field EoS parameter becomes roughly $-\frac{1}{2}$ in the presence of 
matter and remains unchanged up to $N \sim -1$, after which it evolves to $-1$ for late time. After radiation-matter equality, the total EoS parameter becomes zero representing the deceleration in 
expansion up to $N\sim -3$ , after which it starts to drop
to $-1$ referring to late time accelerated expansion. The transition from the deceleration to acceleration in 
expansion happens a bit later in our model compared to the $\Lambda$CDM.
The late time predictions for $w_g$ and $w_{tot}$ are in good agreement with those of Ref. \cite{cai11} for constant $\alpha$ and non-interacting case.

%The extension of the model toward the early universe shows us that early time model predictions do not spoil the predictions of the standard Big Bang model and we also learn interesting properties of the ghost field at higher temperatures, such as the dynamical EoS parameter. 

In the case of Hubble parameter, our prediction is consistent with that of the $\Lambda$CDM  and better fits to the recent observational data at late time compared to the prediction of Ref. \cite{urban10}
for constant  $\alpha$ existing in the interval $-1.2<N<-0.2$. All models have the same predictions at $N=0$. 

 For the deceleration parameter, our model predicts that the change in sign of this parameter (transition from the decelerated expansionary to the accelerated expansionary phase) 
occurs a bit later than that of the $\Lambda$CDM model although these models have roughly the same predictions on the behavior of $q$ with respect to $N$ at late time.

The obtained results at finite temperature  in the present work  point out that the QCD ghost vacuum can be  still a valid candidate to dark energy. 
Interestingly, this dark energy source has a dynamical equation of state parameter equal to $-1/3$ at the early universe and $-1$ today, behaving similar 
to the cosmological constant at late time. This dynamical property of EoS parameter  can be checked by fitting to  observations.

\begin{acknowledgments}
The authors thank A. R. Zhitnitsky, M. Ar{\i}k and  \" O. Akarsu for useful discussions. N. Kat{\i}rc{\i} thanks Bo\u{g}azi\c{c}i university for the financial support provided through the scientific research (BAP) project with grant no $7128$.
\end{acknowledgments}

\end{document}